# The Proper Analytical Solution of the Korteweg-de Vries-Burgers Equation

Jian-Jun SHU

*Abstract*—The asymptotic expansion of the Korteweg-de Vries-Burgers equation is presented in this paper.

## I. Introduction

It is common knowledge that many physical problems (such as non-linear shallow-water waves and wave motion in plasmas) can be described by the Korteweg-de Vries (KdV) equation [1]. It is well known that solitons and solitary waves are the class of special solutions of the KdV equation. In order to study the problems of liquid flow containing gas bubbles [2], fluid flow in elastic tubes [3], and so on, the governing equation can be reduced to the so-called Korteweg-de Vries-Burgers (KdVB) equation as follows [4]:

$$\frac{\partial u}{\partial t} + u \frac{\partial u}{\partial x} - \alpha \frac{\partial^2 u}{\partial x^2} + \beta \frac{\partial^3 u}{\partial x^3} = 0. \quad (1)$$

This equation is equivalent to the KdV equation if a viscous dissipation term ($\alpha \frac{\partial^2 u}{\partial x^2}$) is added. The studies of the KdV equation [1] ($\alpha = 0$) and the Burgers equation [5] ($\beta = 0$) have been undertaken, but the exact solution for the general case of equation (1) ($\alpha > 0$, $\beta > 0$) has still not been completed.

In studying the theory of ordinary differential equations and applications, it is clear that the asymptotic expansion is highly important. The asymptotic expansion of equation (1) is presented in this paper. It would be useful to understand thoroughly the property of the solution to the KdVB equation. The asymptotic expansion would provide a reliable basis for estimating the advantages, and disadvantages of numerical methods associated with equation (1).

## II. Transformation

The following new variable is introduced as [6], [7]:

$$\zeta = x - \lambda t \quad (2)$$

Equation (1) can be written as

$$\frac{\partial u}{\partial t} + (u - \lambda)\frac{\partial u}{\partial \zeta} - \alpha \frac{\partial^2 u}{\partial \zeta^2} + \beta \frac{\partial^3 u}{\partial \zeta^3} = 0. \quad (3)$$

The so-called travelling-wave solution, i.e. $u = f(\zeta)$, shall be considered here. By integrating formula (3) with respect to $\zeta$, a non-linear differential equation can be obtained as follows:

$$u_{\zeta\zeta} + c_1 u_\zeta + c_2 u^2 + c_3 u = c_0 \quad (4)$$

where $c_1 = -\frac{\alpha}{\beta}$, $c_2 = \frac{1}{2\beta}$, $c_3 = -\frac{\lambda}{\beta}$ and the integral constant $c_0 > -\frac{\lambda^2}{2\beta}$. If $c_0 \neq 0$, a simple translation transformation, $u = \tilde{u} + \tilde{c}_0$ ( $\tilde{c}_0 = \frac{-c_3 \pm \sqrt{c_3^2 + 4 c_0 c_2}}{2 c_2}$ ), can be made. $\tilde{u}$ satisfies the following equation: $\tilde{u}_{\zeta\zeta} + c_1 \tilde{u}_\zeta + c_2 \tilde{u}^2 + (c_3 + 2 c_2 \tilde{c}_0)\tilde{u} = 0$. Without loss of generality, we shall confine ourselves to the consideration of $c_0 = 0$ alone from now on. It can be further assumed that $\lambda \geq 0$, because the discussion on $\tilde{\lambda} = -\lambda$ can be made in the same manner for $\lambda < 0$. Equation (4) can be written as a system of first-order equations:

$$\frac{du}{d\zeta} = v \quad (5)$$
$$\frac{dv}{d\zeta} = -c_2 u(u - 2\lambda) - c_1 v.$$

According to the qualitative theory of ordinary differential equations, the system of first-order equations (5) has two singular points, ($0$, $0$) and ($2\lambda$, $0$). The singular point ($0$, $0$) is invariably a saddle point. The singular point ($2\lambda$, $0$) has three different cases which depend on the values of $\alpha$, $\beta$ and $\lambda$.

A  If $\alpha \geq 2\sqrt{\beta\lambda}$, ($2\lambda$, $0$) is a nodal point.
B  If $0 < \alpha < 2\sqrt{\beta\lambda}$, ($2\lambda$, $0$) is a focal point.
C  If $\alpha = 0$, ($2\lambda$, $0$) is a central point.

The three classes of solutions are roughly shown in Fig. The asymptotic expansion is one which is real and continuous if the argument is greater than a certain value. The following variable transformation can be made:

$$u = -e^{-\frac{c_1(1-k)\zeta}{2}} \frac{c_1^2 k^2}{c_2} y(\xi), \qquad \xi = e^{-c_1 k \zeta} \quad (6)$$

where $k = \sqrt{1 - \frac{4 c_3}{c_1^2}} = \sqrt{1 + \frac{4\beta\lambda}{\alpha^2}} \geq 1$ is a constant. Equation (4) ($c_0 = 0$) can be reduced to the Emden-Fowler equation

$$\frac{d^2 y}{d\xi^2} = \xi^\sigma y^2 \quad (7)$$

where $\sigma = \frac{1 - 5k}{2k}$. It is obvious that $\sigma = -2$ if $\lambda = 0$, and $-2 > \sigma \geq -\frac{5}{2}$ if $\lambda > 0$. Some characteristics of the KdVB equation can be derived based on equation (7).

Jian-Jun SHU is with School of Mechanical & Aerospace Engineering, Nanyang Technological University, 50 Nanyang Avenue, Singapore 639798 (e-mail: mjjshu@ntu.edu.sg).

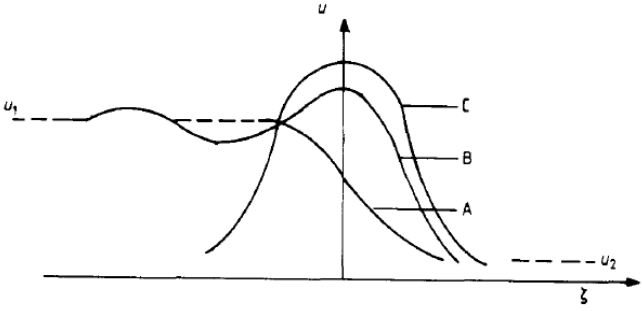

Fig. Typical solutions of the KdVB equation: A is a dissipation-dominant solution for a monotonic shock wave, B is a chromatic dispersion-dominant solution for an oscillatory shock wave, and C is a solitary-wave solution of the KdV equation.

### III. ASYMPTOTIC EXPANSION

*Definition.* Let $y(\xi)$ be an arbitrary function of $\xi$. If $\xi_0$ is a zero point and no other zero point except $\xi_0$ exists in the open interval ($\xi_0 - \varepsilon$, $\xi_0 + \varepsilon$) for some $\varepsilon > 0$, $\xi_0$ is called an isolated zero point of function $y(\xi)$.

*Theorem 1.* The KdVB equation (4) has finite isolated zero points only.

*Proof.* Since $\frac{d^2 y}{d\xi^2} = \xi^\sigma y^2$, $y''$ does not change its sign for $\xi \in (0, +\infty)$. Equation (7) has finite zero points only, except that it identically vanishes for some intervals. Equation (7) has finite zero points only. The theorem is proved. ∎

Theorem 1 indicates that the solution of the KdVB equation is consistently positive, negative or zero for large arguments, which depends upon the condition of infinite point. In order to obtain the asymptotic expansion of the KdVB equation, the following three lemmas are introduced here.

*Lemma 1. Integral rule for asymptotic formulae.* Let $\varphi(t) \sim f(t)$, where $f(t) \neq 0$ and does not change in sign, $\int_{t_0}^{t} \varphi(t) dt \sim \int_{t_0}^{t} f(t) dt$ if $\int_{t_0}^{+\infty} |f(t)| dt = +\infty$ and $\int_{t}^{\infty} \varphi(t) dt \sim \int_{t}^{\infty} f(t) dt$ if $\int_{t_0}^{+\infty} |f(t)| dt < +\infty$.

*Lemma 2. Character of asymptotic expansion.* If $f(t) > 0$, and if $f'$ is a continuous and non-negative function as $t \geq t_0$, $f' \leq f^{1+\varepsilon}$ for $t \geq t_0$ any $\varepsilon > 0$, except perhaps in a set of intervals of finite total length, which depends upon $\varepsilon$.

*Lemma 3. Hardy's theorem.* Any solution of equation $\frac{df}{dt} = \frac{P(f,t)}{Q(f,t)}$, to be continuous for $t \geq t_0$, is ultimately monotonic, together with all its derivatives, and satisfies one of relations, $f \sim at^b e^{E(t)}$ or $f \sim at^b (\ln t)^c$, where $E(t)$ is a polynomial in $t$ and $a$, $b$, $c$ are constants.

The above three lemmas can be adopted to derive the asymptotic expansion of the KdVB equation.

*Theorem 2.* If $\lambda > 0$, the negative asymptotic expansion of the KdVB equation has the following form:
$$u = -\frac{2k^2 v^2 u_\infty}{\delta} e^{-\frac{(k-1)v\zeta}{2\delta}} - \frac{8k^4 v^2 u_\infty^2}{(k-1)(3k-1)\delta} e^{-\frac{(k-1)v\zeta}{\delta}} [1 + O(1)] \quad (8)$$
as $\zeta \to +\infty$, where $k = \sqrt{1 + \frac{4\lambda\delta}{v^2}}$, and $u_\infty > 0$ is constant.

*Proof.* Here $-2 > \sigma > -\frac{5}{2}$ for $\lambda > 0$. If $u$ has a negative asymptotic expansion, $y$ has a positive asymptotic expansion. Since $\frac{d^2 y}{d\xi^2} = \xi^\sigma y^2 > 0$ for $\xi > 0$, $y'$ must be strictly monotonically increasing for $\xi \in (0, +\infty)$ and $y$ must be a monotone function for large $\xi$. Thus $y'$ has three possible cases as $\xi \to +\infty$: (1) $y' \to 0$, (2) $y' \to y_0' = \text{const} > 0$, and (3) $y' \to +\infty$.

Let us show that case (2) is impossible. If $y' \to y_0' = \text{const} > 0$, $y \sim y_0' \xi$, and from equation (7), $y'' = \xi^\sigma y^2 \sim y_0'^2 \xi^{\sigma+2} > \frac{1}{2} y_0'^2 \xi^{\sigma+2}$, whose integration yields $y' > \frac{y_0'^2}{2(\sigma+3)} \xi^{\sigma+3} \to +\infty$ for large $\xi$, which leads to a contradiction.

Then it will be shown that case (3) is impossible. If $y' \to +\infty$, $y' > M$ for large $\xi$ and some $M > 0$, and hence $y > Mx$. Reverting to equation (7), $y'' = \xi^\sigma y^2 > M^2 \xi^{\sigma+2}$, $y > \frac{M^2}{(\sigma+3)(\sigma+4)} \xi^{\sigma+4}$ for large $\xi$. Continuing in this fashion, $y > y_0 \xi^5$ can be obtained for large $\xi$ and the constant $y_0$. Hence from equation (7), $y'' = \xi^\sigma y^2 > \sqrt{y_0} y^{\frac{3}{2}}$ for large $\xi$. Since $y'$ is positive, $y' y'' > \sqrt{y_0} y^{\frac{3}{2}} y'$, whose integration yields $y' > \frac{2 y_0^{\frac{1}{4}}}{\sqrt{5}} y^{\frac{3}{4}}$, which is impossible due to Lemma 2.

Consequently it is left with case (1), where $y' \to 0$. Since $y' < 0$ is strictly monotone increasing for $\xi \in (0, +\infty)$, and $y$ is strictly monotone decreasing for $\xi \in (0, +\infty)$. Since $y > 0$ for large $\xi$, $y$ has a finite limit $u_\infty \geq 0$ as $\xi \to +\infty$.

Let us show that $u_\infty \neq 0$. If $u_\infty = 0$, $y(\xi_0) = \delta > 0$ is set to be small. Since $y$ is strictly monotone decreasing,
$$\delta = y(\xi_0) = \int_{\xi_0}^{+\infty} \left( \int_{t}^{+\infty} \tau^\sigma y^2 d\tau \right) dt < \delta^2 \int_{\xi_0}^{+\infty} \left( \int_{t}^{+\infty} \tau^\sigma d\tau \right) dt$$
or
$$\delta > \frac{(\sigma+1)(\sigma+2)}{\xi_0^{\sigma+2}},$$
which leads to the contradiction for $\delta$ sufficiently small.

Then let $y(+\infty) = u_\infty > 0$, $y(\xi) = u_\infty + O(1)$ as $\xi \to +\infty$:
$$y'(\xi) = -\int_{\xi}^{+\infty} y'' dt = -\int_{\xi}^{+\infty} t^\sigma y^2 dt = \frac{u_\infty^2}{\sigma+1} \xi^{\sigma+1} [1 + O(1)] \quad \text{and thus}$$
$$y(\xi) = u_\infty - \int_{\xi}^{+\infty} y' dt = u_\infty + \frac{u_\infty^2}{(\sigma+1)(\sigma+2)} \xi^{\sigma+2} [1 + O(1)]. \quad \text{The}$$

theorem is proved. ∎

*Theorem 3.* If $\lambda = 0$, the negative asymptotic expansion of the KdVB equation has the following form:

$$u \sim -\frac{2k\nu}{\zeta} e^{-\frac{(k-1)\nu\zeta}{2\delta}} \quad (9)$$

as $\zeta \to +\infty$, where $k = \sqrt{1 + \frac{4\lambda\delta}{\nu^2}}$.

*Proof.* Here $\sigma = -2$ for $\lambda = 0$. If $u$ has a negative asymptotic expansion, $y$ has a positive asymptotic expansion.

Let $\xi = e^s$, obtaining from equation (7),

$$\frac{d^2 y}{ds^2} - \frac{dy}{ds} - y^2 = 0. \quad (10)$$

If $\frac{dy}{ds} = 0$ at $s_0$, $\frac{d^2 y}{ds^2} = y^2 > 0$, and $y$ can only have a minimum at $s_0$. Hence $y$ is a monotone function for large $\xi$. Thus $y$ has three possible cases as $s \to +\infty$: (1) $y \to 0$, (2) $y \to y_0 = \text{const} > 0$, and (3) $y \to +\infty$.

Let us show that case (2) is impossible. If $y \to y_0 > 0$, $\frac{d^2 y}{ds^2} - \frac{dy}{ds} \sim y_0^2$. By integration, $\frac{dy}{ds} - y \sim y_0^2 s$ is obtained. Since $y \to y_0 > 0$, this implies $\frac{dy}{ds} \sim y_0^2 s$, from which $y \sim \frac{1}{2} y_0^2 s^2$, which contradicts $y \to y_0$.

Let us show that case (3) is impossible. If $y \to +\infty$, let $p = \frac{dy}{ds}$, equation (10) becomes

$$p\frac{dp}{dy} - p - y^2 = 0. \quad (11)$$

Since $y \to +\infty$, $p = \frac{dy}{ds} > 0$. Lemma 3 indicates that $p$ has two possible cases for large $y$: (i) $p \sim ay^b e^{E(y)}$, and (ii) $p \sim ay^b (\ln y)^c$, where $E(y)$ is a polynomial in $y$ and $a > 0$, $b$, $c$ are constants.

Let us show that case (i) is impossible. If $E(y) \to -\infty$, $p \to 0$ and $\frac{dp}{dy} \to 0$, which lead to a contradiction by referring to equation (11). If $E(y) \to +\infty$, $p > y^2$ for large $y$, which is impossible due to Lemma 2. Hence $E(y) = \text{const}$.

If $b > 1$, $p > y^{\frac{b+1}{2}}$ for large $y$, which is impossible due to Lemma 2. If $b \le 1$, $p\frac{dp}{dy} \sim y^2$ is obtained from equation (11). By integration, $\frac{1}{2} p^2 \sim \frac{1}{3} y^3$ is obtained, so that $b = \frac{3}{2} > 1$, which leads to a contradiction.

Let us show that case (ii) is impossible. If $b > 1$, $p > y^{\frac{b+1}{2}}$ for large $y$, which is impossible due to Lemma 2. If $b \le 1$, $p\frac{dp}{dy} \sim y^2$ is obtained from equation (11). By Integration, $\frac{1}{2} p^2 \sim \frac{1}{3} y^3$ is obtained, so that $b = \frac{3}{2} > 1$, which leads to a contradiction.

Consequently it is left with case (1), where $y \to 0$. Let $v = \frac{1}{y}$ and $w = \frac{dv}{ds}$, obtaining from equation (10),

$$w\frac{dw}{dv} - \frac{2w^2}{v} - w + 1 = 0. \quad (12)$$

Since $y \to 0$, $v \to +\infty$ and $\frac{dy}{ds} < 0$, $w = \frac{dv}{ds} = -\frac{1}{y^2}\frac{dy}{ds} > 0$ is obtained. Lemma 3 indicates that $w$ has two possible cases for large $v$: (iii) $w \sim av^b e^{E(v)}$, and (iv) $w \sim av^b (\ln v)^c$, where $E(v)$ is a polynomial in $v$ and $a > 0$, $b$, $c$ are constants.

It is now shown that if case (iii) is satisfied, $E(v) = \text{const}$ and $b = 0$.

Similar to above, $E(v) = \text{const}$ and $b \le 1$.

If $b = 1$, $\frac{dw}{dv} \sim a > 0$. From equation (12), $a = -1$ is obtained, which leads to a contradiction.

If $0 < b < 1$, $\frac{dw}{dv} \sim 1$ is obtained from equation (12). By integration, $w \sim v$ is obtained, so that $b = 1$, which leads to a contradiction.

If $b < 0$, $w\frac{dw}{dv} \sim -1$ is obtained from equation (12). By integration, $\frac{1}{2} w^2 \sim -v$ is obtained, which leads to a contradiction.

Let us show that if case (iv) is satisfied, $b = 0$ and $c = 0$.

Similar to above, either $b = 1$, $c \ne 0$, or $b = 0$.

If $b = 1$, $c < 0$ or $c > 0$, $\frac{dw}{dv} \sim 1$ or $\frac{dw}{dv} \sim \frac{2w}{v}$ is obtained from equation (12). By integration, $w \sim v$ or $w \sim v^2$ is obtained, so that $c = 0$, which leads to a contradiction. Hence $b = 0$.

If $c < 0$, $w\frac{dw}{dv} \sim -1$ is obtained from equation (12). By integration, $\frac{1}{2} w^2 \sim -v$ is obtained, which leads to a contradiction.

If $c > 0$, $\frac{dw}{dv} \sim 1$ is obtained from equation (12). By integration, $w \sim v$ is obtained, so that $c = 0$, which leads to a contradiction.

Summing up and from equation (12), $w \sim 1$ is obtained, so that $\frac{dv}{ds} \sim 1$ as $s \to +\infty$. By integration, $v \sim s$ is obtained as $s \to +\infty$, so that $y \sim \frac{1}{\ln \xi}$ as $\xi \to +\infty$. The theorem is proved. ∎

*Theorem 4.* If $\lambda > 0$, the negative asymptotic expansion of the KdVB equation can be written as

$$u = -\frac{2k^2 \nu^2 u_\infty}{\delta} e^{-\frac{(k-1)\nu\zeta}{2\delta}} - \frac{2k^4 \nu^2}{\delta} \sum_{i=1}^{+\infty} \frac{(2u_\infty)^{i+1} e^{-\frac{(i+1)(k-1)\nu\zeta}{2\delta}}}{\prod_{j=1}^{i}[j(k-1) + 2k]j(k-1)} \quad (13)$$

where $k = \sqrt{1 + \frac{4\lambda\delta}{\nu^2}}$ and $u_\infty > 0$ is constant.

*Proof.* Since $e^{-\frac{(i+1)(k-1)\nu\zeta}{2\delta}}$ exists, the infinite series converges. Let

$$u_m = -\frac{2k^2v^2 u_\infty}{\delta} e^{-\frac{(k-1)v\zeta}{2\delta}} - \frac{2k^4 v^2}{\delta} \sum_{i=1}^{m} \frac{(2u_\infty)^{i+1} e^{-\frac{(i+1)(k-1)v\zeta}{2\delta}}}{\prod_{j=1}^{i}[j(k-1)+2k]j(k-1)},$$

$$y_m = u_\infty + \sum_{i=1}^{m} \frac{2^{i-1} u_\infty^{i+1} \xi^{i(\sigma+2)}}{\prod_{j=1}^{i}[j(\sigma+2)-1]j(\sigma+2)}.$$

$$y_{m+1}[1+O(1)] = u_\infty + \int_\xi^{+\infty}\left(\int_t^{+\infty} \tau^\sigma y_m^2 [1+O(1)]^2 d\tau\right)dt \quad \text{can be}$$

obtained for an arbitrary integer $m$. Since $u_m \to u$ as $m \to +\infty$, $y_m \to y_{+\infty}$ as $m \to +\infty$, so that $y_{+\infty} = u_\infty + \int_\xi^{+\infty}\left(\int_t^{+\infty} \tau^\sigma y_{+\infty}^2 d\tau\right)dt$ and $y_{+\infty}$ is the positive asymptotic expansion of equation (7). The theorem is proved. ∎

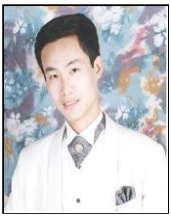

**Jian-Jun SHU** is a recipient of the British Institution of Mechanical Engineers 1992 BFPA Prize for Young Engineers. He is a member of the editorial board of the journal of Mathematical Problems in Engineering. He has published over 80 technical papers and presented over 40 invited lectures/seminars. He is with Nanyang Technological University, Singapore.